\documentclass{article} 
\begin{document} 
\begin{center} 
\title{Real Observers and the Holographic Principle} 
\author{M.C.Dance} 
\end{center} 
\begin{abstract} 

The holographic principle asserts that the observable number of degrees of freedom inside a volume is proportional not to the volume, but to the surface area bounding the volume. There is currently a need to explain the principle in terms of a more fundamental microscopic theory. This paper suggests a potential explanation. This paper suggests that in general, for an observer to observe the $r$ coordinate of an event, the process of making that observation must generate at least as much entropy as the information that the observation gains. Following on from that, this paper sets out a simple argument that leads to the result that observers on the surface of a sphere can observe an amount of information about the enclosed system that is no more than an amount that is proportional to the surface area of the sphere. 
\end{abstract} 

\section{Introduction} 

The holographic principle asserts that the observable number of degrees of freedom inside a volume is proportional not to the volume, but to the surface area bounding the volume. A summary of the principle, its history and applications has been provided by Bousso~\cite{bousso_2001}. The principle envisages a set of observers located on the surface in question. The principle arose originally from Bekenstein's suggestion that the entropy of a black hole is proportional to the area of the hole's horizon. It has since been generalized by Susskind to other systems, and examples of existing field theories that exemplify the principle have been widely discussed in the literature.

There is a need to explain the principle in terms of a more fundamental microscopic theory. 

\section{Summary}

In this paper, I make a suggestion regarding the abilities of observers. The suggestion may provide a clue to the apparent dimensional reduction that the holographic principle implies. 

The suggestion is that in general, "intelligence" is required in order for an observer located (say) around the origin to observe the $r$ coordinate of an event. If this is so, and if an intelligent observation generates at least as much entropy as the information it gains, as originally suggested in 1929 by Szilard~\cite{szilard_1929}, the holographic principle may be a consequence. Strictly speaking, Szilard's result was derived for a classical measurement. However, Zurek~\cite{zurek_2003} in 2003 showed that an "intelligent" quantum mechanical observation also must generate at least as much entropy as the information it gains. These results, for classical and quantum mechanical measurements, are underpinned by the requirement that the second law of thermodynamics must hold. 

If we assume, then, that an observer must be "intelligent" in order to measure $r$, the argument proceeds as follows. If the entropy produced by the observer reduces the information that the observer has about the observed system by an amount equal to the produced entropy, while the produced entropy is greater than or equal to the information gained by the observer from measuring the $r$ of an event in the system, then an $r$ measurement gives the observer no additional information about the state of the observed system. The observer would in principle then be limited to an amount of information about the observed system that would be governed by the remaining (non-$r$) spatial degrees of freedom of the observed system, at any given time. 

The final step in the reasoning notes that this number of degrees of freedom is proportional to the surface area of the observed system's boundary, for a spherical observed system, due to the minimum observable distance interval that founds non-commutative geometry. A description of this step in equation form appears below.

One might think that a difficulty with this approach is that entropy is linked to the number and accessibility of the microscopic states of the system. If an observation increases the entropy of the system, or keeps the entropy constant, that implies that the number of states accessible to the system has increased, or at least has remained the same. One might think that this would achieve the opposite of the drastic reduction in observable degrees of freedom that the holographic principle asserts. 

This difficulty can possibly be resolved by realizing that the key word in the holographic principle (as expressed above) may be "observable". In this paper, we are concerned with how much information about the observed system is accessible to the observer. If as a result of an observation, entropy is added to the system so that more states become accessible to the observed system, the observer has lost the corresponding amount of information about the observed system.

The following sections describe the argument in more detail.

\section{The Observer} 

It is well known that Heisenberg's uncertainty principle~\cite{hberg_1927}, when applied to an observer, implies that there are non-zero minimum observable distance and time intervals that depend on the observer's mass. 

I conjecture that further properties of an observer (ie other than the finiteness of its mass) further restrict what it can observe. One of these properties is presence or otherwise of "intelligence". An intelligent observer may be able to perform calculations in order to compute a coordinate of an event. An intelligent observer must be an open thermodynamic system.

In my view, all observations at the microscopic (quantum) level are carried out at the initial stage by quantum mechanical observers. Such a quantum observer might be a single atom. A large-scale observer is a composite of microscopic quantum mechanical observers.

An example of an intelligent observer is the retina. The initial stage of photon detection (not in itself obviously intelligent) in the retina involves an alteration in the conformation of a single covalent bond within a single retinal molecule. The conformation change is an inherently quantum mechanical event. Later stages of visual processing amplify and filter the initial quantum mechanical observation event in a way that requires a supply of energy to the processing cells. It is intelligent processing and requires that the observer is an open thermodynamic system.

\section{Observing the Polar Coordinate $r$ of an Event} 

Let us consider an event E that occurs at $(t,r,\theta,\phi)$ (classically speaking), and an observer O at (near) $(0,0,0,0)$. Suppose that particles are generated at E and they arrive at O. Can O detect $r$?

\subsection{Simple Observer} 

Let us first consider a simple quantum mechanical observer, such as an atom, that simply detects particles arriving. This observer has no intelligence, and does not plan observations. 

It seems that O might (or might not) be able to inherently detect $\theta$ and/or $\phi$. For example, if O is axially symmetric about an internal $z$ axis, the manner in which O detects the particle may produce information about $\theta$. 

I conjecture that O cannot detect $r$ directly. This is because I cannot presently imagine any manner in which O's detection of the particle can provide information on the $r$ at which the event occurred, without O making some kind of calculation. As far as O can tell from the measurements (without more), the particles could have been generated at any points along any possible paths of travel leading to their detection, at any distance away from the detector. Even from a classical trajectory viewpoint, E could be anywhere along the line that extends from the point of detection at the origin and that is defined by $(\theta , \phi )$ (assuming for now that $\theta$ and $\phi$ have been observed). 

The same would be true for an observer that is an extended molecule. Even then, a single incoming particle would generally be absorbed in a way which would not provide any information about $r$.

Can the simple observer measure $r$ indirectly? Suppose that $r=vt_0$, where the particle was emitted at $t_0$ with speed $v$. To observe $r$, the observer must measure both $v$ and $t_0$, and must compute $r$. This appears to ask too much of a simple quantum observer such as an atom. 

It therefore appears that the simple quantum observer generally does not observe $r$ for events. In a space-time with three spatial dimensions, it seems that the simple quantum mechanical observer generally can observe at most two spatial coordinates: $\theta$ and $\phi$. Supposing for a moment that there is such an observer that can observe $\theta$ and $\phi$, let us call such an observer a 2D-observer.

\subsection{Complex Observer} 

When we think about distance measurements, we generally think of an observer who plans the measurement. The observer emits a probe particle, the probe particle reflects off an event E, and finally the observer receives the probe particle. Provided that the observer has a clock and knows the particle's initial and final speeds, the observer can estimate the distance r to E. This observer is clearly an "intelligent" system.

Alternatively, the observer might have set up a frame of reference (in a classical picture anyway), with rulers and clocks. An individual clock or point on a ruler cannot infer much about the world around it. However, the (higher level) observer takes readings of various clocks and rulers and works out where and when E must have happened from those readings. I conjecture that such an observer is "intelligent". 

\section{Holographic Principle} 

\subsection{Single Simple Observer at Origin of Spherical Observed System} 

According to ideas suggested above, a simple 2D-observer sitting at the centre of a three-dimensional spherical volume in space might observe at most $\theta$ and $\phi$ information arriving from events in the volume. The observer does not detect the $r$ coordinates of the events. The question then becomes: how many $(\theta, \phi)$ combinations can the observer distinguish? The number of distinguishable combinations will be equal to an upper limit on the observable number of degrees of freedom.

Suppose the sphere has radius $R$. (Our simple 2D-observer does not know this, but we do!) Then the spatial separation of two closely adjacent points $(\theta_1 , \phi_1)$ and $(\theta_2, phi_2)$ on the surface of the sphere is approximately $R(|\theta_1-\theta_2|+|\phi_1-\phi_2|)$. 

For our observer to detect these as separate points, the separation must be greater than the minimum observable interval $\Delta d$. For any particular observer, $\Delta d$ depends on the observer's mass and also goes as the cube root of the distance $R$ (shown by Wigner). That said, $\Delta d$ also must be greater than or equal to an absolute minimum value $\Delta d_0$ which depends on the observer's mass but does not depend on $R$.

This defines the observer's $\theta $- and $\phi $-resolution $\Delta \theta $ and $\Delta \phi $:- 

\begin{equation} 
\Delta \theta = \Delta \phi \approx \Delta d / R. 
\end{equation} 

The number of spacetime points that the observer can distinguish is then at most 

\begin{equation} 
N = \frac{4\pi }{\Delta \theta \Delta \phi } 
\end{equation} 

which is approximately equal to $ 4\pi R^2 / ( \Delta d)^2 $.  This expression must be less than or equal to $ 4\pi R^2 / (\Delta d_0)^2 $, because $\Delta d$ must be greater than or equal to the minimum observable distance $\Delta d_0$.

That is, the single observer at the origin can access only an amount of information, about the observed system in the volume, that is less than proportional to the surface area $4\pi R^2$. 

\subsection{Ensemble of Simple Observers at the Surface of an Observed Sphere} 

The holographic principle concerns an ensemble of observers on the surface, rather than a single observer within the volume. 

Therefore, let us again suppose that the observed system is a sphere of radius $R$, but that we now have an ensemble of simple observers, each located just inside the surface of the sphere. We will assume that each observer has the same mass and that all $\theta$ and $\phi$ values can be detected (subject to the minimum observable spatial interval).

Each observer has a maximum spatial resolution limited by $\Delta d_0$. No observer by itself can distinguish between different values of $r$. Then it seems that the total number of degrees of freedom of the observed system that the ensemble of observers (acting independently) can possibly distinguish between (and therefore detect) must be the largest integer that is less than or equal to $4\pi R^2/ (\Delta d_0)^2$.

\subsection{Single Complex Observer Near Origin of Spherical Observed System} 

As discussed above, this paper conjectures that only an intelligent observer can detect the $r$ coordinate of an event. One might think that this intelligence must increase the amount of observable information that the observer can have about the observed system. 

However, it was suggested in 1929 by Szilard~\cite{szilard_1929}that an intelligent observation must generate at least as much entropy as the information obtained by the observation. Interestingly, this 1929 paper by Szilard is reproduced in a classic textbook titled "Quantum Theory and Measurement". This book is a collection of seminal papers in quantum mechanics, and includes the original quantum mechanics paper of Heisenberg~\cite{hberg_1927}, in which Heisenberg pointed out the difficulty that quantum mechanics poses for observers in Einstein's classical theory of relativity. . Strictly speaking, Szilard's result was derived for a classical measurement. Zurek~\cite{zurek_2003} in 2003 showed that an "intelligent" quantum mechanical observation also must generate at least as much entropy as the information it gains. These results, for classical and quantum mechanical measurements, are underpinned by the requirement that the second law of thermodynamics must hold.

Thinking then about our single intelligent observer at (or quantum mechanically, somewhere around) the origin, the question is then where does this generated entropy go, after the observation? It either stays within the observer or it goes out into the spacetime, or a combination of these. Possibly the observer may be able to retain entropy for a time, but presumably retention of entropy above a threshold amount would render the observer inoperative. The more usual scenario would be that the entropy would go out into the surrounding spacetime, ie initially into the observed system.

Now, the intelligent observer might or might not know about this entropy that has gone out into the observed system. However, we know about it! Assuming that the entropy produced by the observer reduces the information that the observer has about the observed system by an amount equal to the produced entropy, while the produced entropy is greater than or equal to the information gained by the observer from measuring the $r$ of an event in the system, the observer could not in principle obtain an amount of information about the observed system that is larger than an upper limit. 

I conjecture that the upper limit would be less than or equal to the remaining (non-$r$) spatial degrees of freedom of the observed system (considering a constant-time snapshot of the system).

To ascertain the upper limit, let us consider a clever observer who has observed every observable $(\theta,\phi)$ combination, without having bothered to measure any $r$ values. The observer has measured the maximum amount of information that might be accessible to a simple observer, and for the purposes of working out an upper limit we shall assume that the observer has generated no entropy in the process of making these measurements. Now let our clever observer measure the $r$ value of an event that has occurred within the volume. 

Assuming Szilard was right, I would conjecture that the observer cannot be {\em a priori} certain that it has any more information about the present state of the system now, than the observer had about the former state of the observed system before it made the $r$ measurement. Moreover, I conjecture that if the observer tries to work out somehow whether it does in fact have more information about the system than it had before, the process of working that out would create at least as much disorder, to negate the information that had been ostensibly gained. 

Therefore, the cautious observer might proceed on the basis that it can get no more information about the observed system inside the sphere than the simple observer can. That is, the observer at the centre of the sphere can reliably access only an amount of information about the system that is at most proportional to the surface area of the sphere.

For a collection of complex observers located on the surface of a sphere, one might use an argument similar to the above to arrive at a similar result. That is, one would expect that the maximum information that such a set of observers can acquire would be again at most proportional to the surface area of the sphere. Such an argument is presented below.

\subsection{Ensemble of Complex Observers On (Near) Surface of Observed Spherical System} 

Now considering an ensemble of intelligent observers at (or quantum mechanically, somewhere around) the surface of a sphere which is the observed system, the question is then where does this generated entropy go, after an $r$ observation by one of the observers? As for the single intelligent observer at the origin, the entropy either stays within the observer or it goes out into spacetime, or a combination of these. Again, the observer may be able to retain entropy for a time, but presumably retention of entropy above a threshold amount would render the observer inoperative. Let us assume then that all of the entropy goes out into the surrounding spacetime.

Because the observer concerned is not at the centre of the observed system, but is instead on its surface now, it is no longer the case that all of the observer-generated entropy must go into the observed system. In general, some of the entropy will go into the observed sphere, and some of it will go outside the sphere. One might think then that the observer will generally have gained information about the system by making the $r$ measurement. While this is true, is it enough to enable the observer (or ensemble of observers) to give us enough information to formulate a physical theory for the observed system that contains more degrees of freedom than the upper limit described above for the single observer at the origin? 

I conjecture that such an on-average information gain is not enough. To formulate a physical theory, we need reliable data. That is, our starting point is observed information on which we know we can rely with (almost) 100 percent certainty. 

I conjecture that the observer on the surface cannot be {\em a priori} certain that it has any more information about the overall state of the system after the $r$ measurement, than the observer had about the overall state of the observed system before the observer made the $r$ measurement. I conjecture this because, even if the intelligent observer knows about the entropy it has generated, the observer has no way of knowing how much of that entropy has gone into the observed system and how much has gone outside the system, unless the observer performs another measurement to work that out. This measurement itself would generate at least as much entropy as the information that this measurement would gain, and the observer does not know where {\em this} entropy has gone. And so on {\em ad infinitum,} so that if the observer tries to work out whether its initial $r$ measurement actually gave the observer any more information about the system than the observer had before, the process of working that out would generate at least as much disorder, to negate the information gained. 

If the reasoning above is correct, it would mean that the ensemble of observers can get no more reliable information about an observed system inside the sphere than the maximum information observable by an ensemble of simple observers. That is, the ensemble of complex observers on the surface of the sphere could reliably access only an amount of information about the observed system that is proportional to the surface area of the sphere.

\section{Conclusion} 

This paper has suggested that in general, "intelligence" is required in order to observe the $r$ coordinate of an event. If this is so, and if an intelligent observation also generates at least as much entropy as the information it gains, as suggested in 1929 by Szilard, this effect may limit the amount of information that an observer can possibly be certain of accumulating about an observed system to an amount of information that is proportional to the surface area of a sphere bounding the volume of the observed system. This looks rather similar to the holographic principle.


\begin{thebibliography}{4}
\bibitem{bousso_2001}R. Bousso, hep-th/0203101. 
\bibitem{hberg_1927}W. Heisenberg, {\em Zeitschrift fur Physik}, 43, 172-98 (1927). English translation reproduced in "Quantum Theory and Measurement", ed. J. Wheeler and W.H. Zurek, Princeton University Press, 1983.
\bibitem{szilard_1929}L. Szilard, "On the Decrease of Entropy in a Thermodynamic System by the Intervention of Intelligent Beings", originally published in {\em Zeitschrift fur Physik} 53, 840-856 (1929); an English translation appears in the collection of papers published under the title "Quantum Theory and Measurement", ed. J.~Wheeler and W.~H.~Zurek, Princeton University Press, 1983. That English translation was originally published in {\em Behavioral Science}, 9, 301-10 (1964).
\bibitem{zurek_2003}W. H. Zurek, "Maxwell's Demon, Szilard's Engine and Quantum Measurements", quant-ph/0301076.
\end{thebibliography}
\end{document}